\documentclass[12pt]{iopart}
\usepackage{iopams}
\usepackage{graphicx}

\begin{document}

\title{Hydrogen adsorption and diffusion around Si(001)/Si(110)
  corners in nanostructures}
\author{Richard Smith$^{1,2,3}$, Veronika Br\'{a}zdov\'{a}$^{1,2,3}$ and David
  R. Bowler$^{1,2,3,4}$}
\address{$^{1}$London Centre for Nanotechnology, UCL, 17-19 Gordon St,
  London WC1H 0AH, UK}
\address{$^{2}$Department of Physics \& Astronomy, UCL, Gower St, London WC1E
  6BT, UK}
\address{$^{3}$Thomas Young Centre, UCL, Gower St, London WC1E
  6BT, UK}
\address{$^{4}$International Centre for Materials Nanoarchitectonics
  (MANA), National Institute for Materials Science (NIMS), 1-1 Namiki,
Tsukuba, Ibaraki 305-0044, JAPAN}
\eads{david.bowler@ucl.ac.uk}

\date{1 October 2013}
\maketitle

\begin{abstract}
While the diffusion of hydrogen on silicon surfaces has been
relatively well characterised both experimentally and theoretically,
the diffusion around corners between surfaces, as will be found on
nanowires and nanostructures, has not been studied.  Motivated by
nanostructure fabrication by Patterned Atomic Layer Epitaxy (PALE), we present a
density functional theory (DFT) study of the diffusion of hydrogen around the edge
formed by the orthogonal (001) and (110) surfaces in silicon.    We find that the barrier from (001) to (110) is approximately 0.3 eV lower than from (110) to (001), and that it is comparable to diffusion
between rows on the clean surface, with no significant effect on the
hydrogen patterns at growth temperatures used.
\end{abstract}

\section{INTRODUCTION }

Historically, bulk silicon chip fabrication has been a planar process
based on repetitive optical lithography.  The progressive reduction in device
surface area has impaired switching characteristics, leading to the introduction
of three-dimensional structures (e.g. finFETs) that can be realized in
planar processes.  This trend has prompted interest in other ways of
engineering three-dimensional silicon growth \cite{Martin:2014jy} such as
Patterned Atomic Layer Epitaxy (PALE) \cite{Lyding:1994pj,Shen:1995az}.
PALE utilizes chemical vapour deposition (CVD) to create silicon
features without the conventional lithography and masking steps.  In this
paper we use density functional theory (DFT) to consider how hydrogen
diffusion around feature edges affects the growth process.

In PALE a hydrogen-passivated Si(001) substrate can be patterned
by selective removal of the hydrogen atoms using an STM probe in UHV
conditions ~\cite{Lyding:1994pj,Shen:1995az,owen2011patterned}: the
exposed dangling bonds provide adsorption sites for Si fragments
deposited from disilane gas admitted into the STM chamber, a form of
UHV-CVD.  In principle, an atomically-precise three-dimensional
structure of any height could be grown by successive depassivation and
deposition steps.  In practice the quality of the grown layers depends
critically on the effectiveness of the depassivation process, which
tends to fall as growth proceeds.  This leads to surface roughening
and the eventual breakdown of epitaxial growth ~\cite{owen2011patterned}.
    
CVD reactions are complex with multiple pathways through the
adsorption, deposition and desorption
phases~\cite{Owen:1997sh,Owen:1997sj}.  On the clean Si(001) surface,
the disilane molecule dissociates into SiH$_2$ groups with accompanying
atomic hydrogen.  When these groups occur on adjacent surface sites they
form monohydride dimers~\cite{Owen:1997sh,Bowler:2003qq} which grow 
into strings and islands upon further disilane deposition.  The hydrogen is adsorbed
directly on the substrate and slowly desorbs into molecular gas.  The ad-dimers and
atomic hydrogen compete for unreacted substrate sites, whose
availability determines the overall reaction rate.  Conventional low
pressure CVD is conducted at relatively high (around 1000 K)
temperatures and the growth rate is determined by precursor
throughput.  With PALE the ambient temperature cannot exceed 550 K
otherwise the passivating H-layer is destabilized.  At this
temperature, desorption is incomplete and atomic hydrogen remains
bound to the substrate, preventing completion of the first monolayer.
Now the surface is saturated with a variety of Si--H configurations and
further depassivation is necessary.  Completion of the first layer typically
occurs after 3 or 4 depassivation and deposition cycles~\cite{owen2011patterned}.

Surface coverage is also affected by hydrogen diffusion at the
prevailing temperature.  H adatoms have been shown to diffuse along
trenches in the the (011) surface with energy barrier of the order 1
eV~\cite{C1CP20108E}.  Other theoretical calculations on the (001) surface
predict somewhat larger barriers, i.e. 1.6 eV~\cite{PhysRevB.54.14153}
to 1.7 eV~\cite{Bowler:1998zv,Durr:2013dq}, in agreement with
experiment~\cite{PhysRevB.54.14153} .
In this study, we use DFT to study H adatoms on the top and side walls of a
notional PALE nanostructure, a simple pillar with a horizontal (001) growth surface.  The sidewalls are
assumed to have the (110) orientation, a surface of technological interest
due to its high carrier mobility.  We investigate whether diffusion
pathways around the corners would inhibit growth, increasing the
number of depassivation/deposition cycles required per monolayer thus rendering
PALE impractical.   We construct a supercell containing both
the (001) and (110) faces and and seek possible diffusion paths around
the corners between these faces.  Finally we calculate the energy barrier between the
two faces and discuss the likely effect on the PALE growth process. 

\section{METHODS}

\subsection{Computational details}

All calculations used density functional theory~\cite{kohn1999nobel}, as implemented in the Vienna Ab-initio
Simulation Package~\cite{kresse1996software} (VASP version 4.6.34) with the
Perdew-Burke-Enzerhof (PBE) generalised gradient approximation (GGA) exchange-correlation
functional~\cite{PhysRevLett.77.3865}.  The VASP projector-augmented-wave
~\cite{PhysRevB.50.17953,PhysRevB.59.1758} (PAW) potentials for silicon and hydrogen were used.  
The PAW-PBE potential file (POTCAR) for the silicon atom
was dated 5th January 2001 and that for hydrogen 15th June 2001.  We used a 200 eV energy cutoff.  The Brillouin zone was sampled with a Monkhorst-Pack mesh~\cite{monkhorst1976special}.  Gaussian
smearing was applied to fractional occupancies with a width of
0.1 eV. The convergence criterion for forces on atoms was 0.01 eV/\AA\
and for total energy 10$^{-4}$ eV. These parameters yield relative
energies and energy barriers reliable to within $\pm$ 0.01 eV.
We used the FIRE algorithm~\cite{PhysRevLett.97.170201} to optimize the structures.
Structural relaxations for hydrogen adsorption and diffusion pathways used the
quasi-Newton method implemented in VASP.

Transition state search was performed using the climbing image nudged
elastic band method as implemented in the VASP Transition State
Tools~\cite{Henkelman:2000xh,Sheppard:2008dl}
(VTST) code. The climbing image variation of NEB converges rigorously
onto the highest saddle point using just a single intermediate point
and yields accurate barrier energy
values~\cite{Klimes:2010fk}.

\subsection{The supercell}

\begin{figure}
\centering
\includegraphics[width=.5\linewidth]{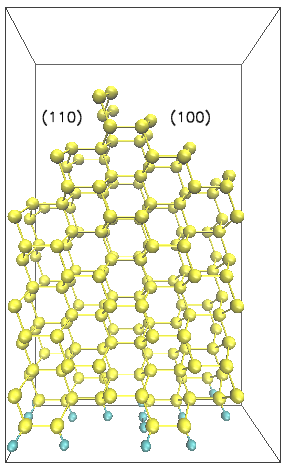}
\caption{The chevron slab with 18$\:$\AA$\:$$\times\:$8$\:$\AA$\;$cross-section and 160 Si atoms.  The Si atoms and H atoms yellow and blue, respectively.  Black lines represent the computational unit cell.}
\label{fig:slab}
\end{figure}

We used a chevron-shaped supercell with a 28$\:$\AA$\:$$\times\:$8$\:$\AA$\;$cross section
with the vertical perpendicular in the (112) direction (Figure~\ref{fig:slab}). This
orientation exposes approximately equal (001) and (110) surface areas
at the apex. A vacuum spacing of 14 \AA\ was interposed and
terminating H atoms added between periodic images.  We used the experimental bulk Si lattice
parameter (5.431 \AA), which is in good agreement with the calculated value for this DFT setup.  The supercell
depth was determined empirically by adding base layers until the
change in energy decrement $\Delta$E was less than 0.01 eV.  The
Monkhorst-Pack sampling grid was established in a similar way and set
to 6 $\times$ 2 $\times$ 1.  The final structure had a depth of 50 \AA\ and contained 350
Si atoms.   The reconstruction of the of the apex region is shown in Figure~\ref{fig:recon} (lower).  We also tested
a smaller supercell with a 18$\:$\AA$\:$$\times\:$8$\:$\AA$\;$cross section and 160 atoms but found that the reconstruction at the
apex showed evidence of strain from the base layers which
disappeared when the structure was enlarged. 

All coordinate files are available freely on figshare~\cite{Smith:2014ys}.

\subsection{Finding H adsorption sites and diffusion pathways}

The H adsorption sites are likely to be 3-coordinate
on the (110) surface or dimers on the (001) surface.
In both cases a single dangling
bond is exposed and its replacement by a covalent Si--H bond causes a
reduction in the total energy. We placed a H atom 1.5 \AA\ above each potential adsorption site on the reconstructed surface (in the direction of the dangling bond) and re-optimized the structure. Total
structure energy was typically reduced by approximately 4  eV when compared to the clean structure.  Binding energies of the adsorbed hydrogen atom were calculated with respect 
to the energy of a gaseous $H_2$ molecule:  

\begin{equation}
E_{\mathrm{b}} = E_{\mathrm{ads}} - E_{\mathrm{clean}} - 1/2\, E_{\mathrm{H}_{2}},
\end{equation}

where $E_{\mathrm{ads}}$, $E_{\mathrm{clean}}$ and $E_{\mathrm{H}_{2}}$ represent the total energy
of the optimised structure with the adsorbed H atom, the optimised energy of
the clean Si reconstruction and the total energy of gas-phase hydrogen molecule, respectively.  A
negative binding energy indicates an energy gain on adsoprtion; a
positive binding energy indicates an energy loss.  A diffusion
pathway will be formed between two adjacent adsorption sites provided that
the energy required to surmount any intervening energy barriers is comparable
with the thermal energy acquired by the mobile atom.  We select the
lowest energy sites close to the step edge and use the NEB climbing-image
optimizer to find the diffusion barriers.  We then use the Arrhenius equation to estimate
the diffusion rate across the edge.

\section{RESULTS AND DISCUSSION}

\subsection{Reconstruction and characterization}

\begin{figure}
\centering
\includegraphics[width=.5\linewidth]{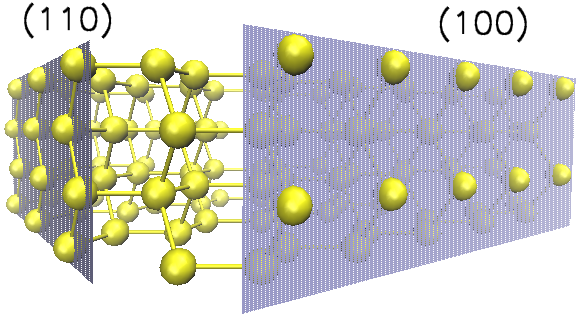}
\includegraphics[width=.5\linewidth]{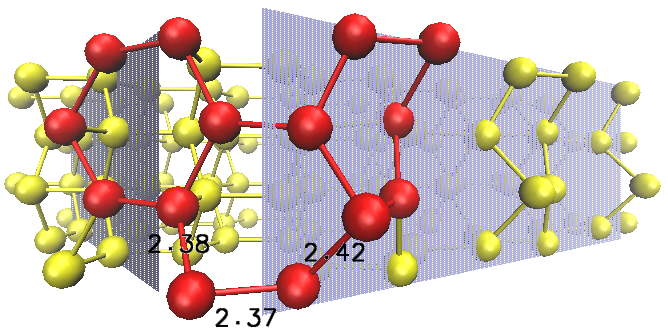}
\caption{(Upper) The apex of the 350-atom unoptimized chevron slab viewed from the (112) direction, showing the (100) and (110) surfaces.  (Lower) Optimized structure, showing (red) buckled (111)-like hexagonal patterns with extended bond lengths (\AA).  The bond length in bulk Si is 2.35 \AA.}
\label{fig:recon}
\end{figure}

As the (001) surface normal swings through a right-angle to the (110)
direction it passes through a number of intermediate surface planes
e.g. [114], [113], [111], [331]  ~\cite{Battaglia:2009km}. Each of
these surfaces has its own reconstruction strategy but all contain the prototypical
3-coordinate surface atom found on the bulk-truncated (111) surface.  Consequently we can expect the reconstruction to
of the apex region to include the hexagonal pattern seen on the (111)
unreconstructed surface.  This can be seen in Figure~\ref{fig:recon}
(upper), while the relaxed structure can be seen in
Figure~\ref{fig:recon} (lower), which also shows the extended bond lengths expected in the
presence of delocalized electrons.  Away from the apex, the figure
shows characteristic dimerization and buckling on the (001) surface
and out-of-plane buckling of the zig-zag rows of the (110) surface.
Distortion in the bulk structure is greatest in the region beneath the
apex and extends to a depth of 20 \AA. These observations suggest that
the reconstruction is plausible, offering a sensible basis for
the calculation of a model potential energy surface. An exhaustive characterization of the step edge
reconstruction would require examination of larger structures in
multiple orientations and is beyond the scope of this work.

\subsection{The potential energy surface}

\begin{figure}
\centering
\includegraphics[width=.5\linewidth]{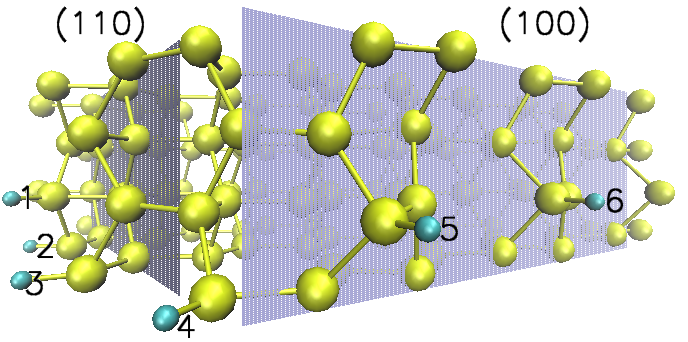}
\includegraphics[width=.5\linewidth]{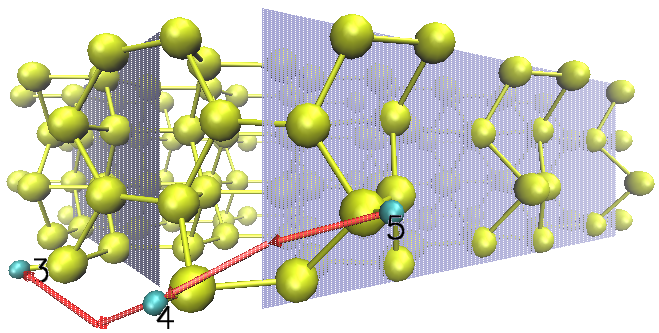}
\caption{(Upper) A composite representation of the 6 most stable absorption sites from 16 surveyed.   Si atoms and H atoms yellow and blue, respectively.  (Lower) Approximate path followed by diffusing H atom (red) in NEB climbing image simulations. } \label{fig:sites} \end{figure}

We investigated 16 possible adsorption sites on the reconstructed surfaces.
Six of these gave total energies falling within a 0.3 eV range while
the remainder was at least 1 eV higher and not considered further further (an increment of 1 eV reduces the adsorption probability by a factor of $\approx 10^4$). These values are shown in Table~\ref{tab:energy} and the sites depicted in Figure~\ref{fig:sites}. 

\begin{table}[here]
\begin{tabular}{lll}
Site & Relative energy & Binding energy \\ 
1 & 0.29 & -1.65 \\ 
2 & 0.06 & -1.88 \\ 
3 & 0.00 & -1.95 \\
4 & 0.02 & -1.92 \\ 
5 & 0.29 & -1.65 \\ 
6 & 0.13 & -1.81 %
\end{tabular}

\caption{Stability of H adsorption sites on the chevron surface. All values are in eV.  Site
  labels are those shown in Figure~\ref{fig:sites} (upper).}
\label{tab:energy}
\end{table}

Sites 3, 4 and 5 are situated on the chevron apex so we considered a diffusion path consisting of the two hops 3--4 and 4--5.  Each is analogous to the kinetics of a chemical reaction with a single transition state corresponding to the highest saddle point in the potential energy landscape lying between the end points.  The climbing image variation of NEB returns the energy at the highest saddle point and so a single-image NEB calculation per hop suffices in this case.  Additional images can provide further points on the reaction path corresponding to the route taken by diffusing atoms, although the computational cost is considerable.  Since we are
interested only in relative barrier heights, we elect to perform single-image calculations. 

\begin{table}
\begin{tabular}{llll}
End-point  & Energy & $\Delta E_{(011\rightarrow100)}$  & $\Delta E_{(100\rightarrow110)}$\\ 
3 & -1947.01 \\
\emph{barrier} & -1945.44 & 1.57 & 1.55 \\
4 & -1946.98 \\ 
\emph{barrier} & -1945.00 & 1.99  & 1.72 \\
5 & -1946.71  %
\end{tabular}

\caption{End-point and NEB barrier energies for a two-hop diffusion pathway around the chevron apex.  All values are in eV.  Site
  labels are those shown in Figure~\ref{fig:sites} (lower).  This data is shown graphically in  Figure~\ref{fig:barriers}.}
\label{tab:barrier}
\end{table}
 
\begin{figure}
\centering
\includegraphics[width=.5\linewidth]{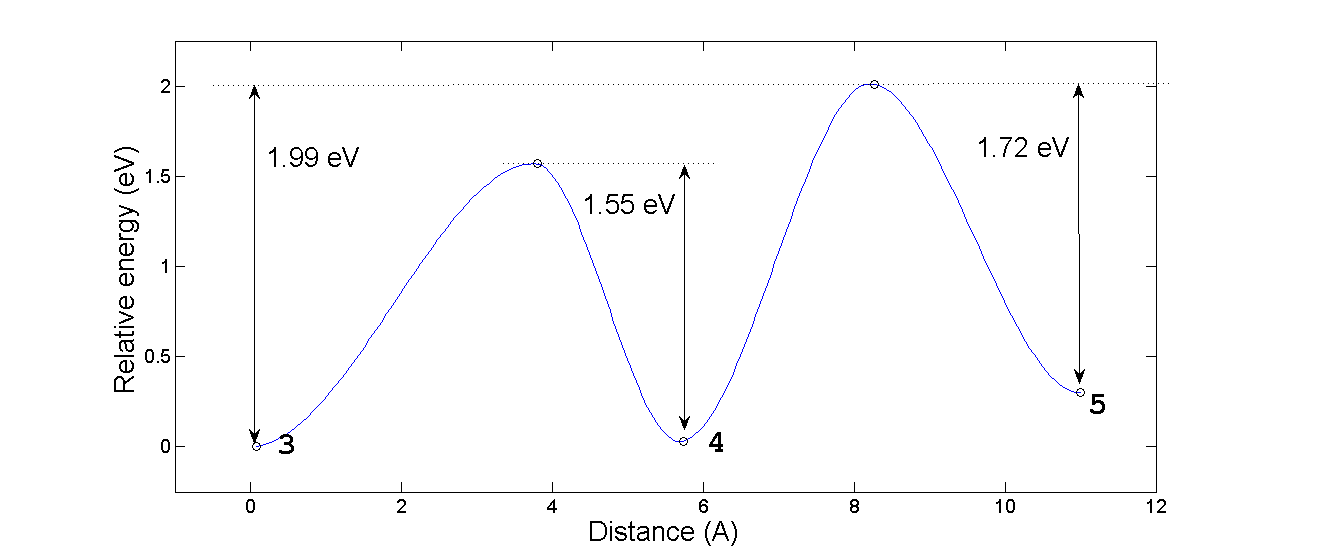}
\caption{Diffusion barriers for a hydrogen atom traversing the PES in the apex region of the chevron slab.   The adsorption sites are labelled as in Fig. 3 and the barrier energies derived from two NEB climbing-image calculations.  The curve is a spline fit to the five data points.  Zero energy at adsorption site 3 corresponds to a calculated value of -1947.01 eV.  Distance travelled increases in the $(110)\rightarrow(100)$ direction.} \label{fig:barriers} \end{figure}

These results of the NEB calculations values are shown in Table~\ref{tab:barrier} and represented graphically in Figure~\ref{fig:barriers}.  Figure~\ref{fig:sites} (lower) gives a rough indication of the actual diffusion path. 
The effective barrier for the path is the greater of the hop barriers and is asymmetrical, due to the differing
starting energies.  Diffusion from (001) to (110) has a barrier of
1.72 eV, while the reverse process has a barrier of 1.99 eV. These can be compared with 
published values of 1.66 eV $\pm$\ 0.15 eV~\cite{Bowler:1998zv} and
1.75 $\pm$\ 0.02 eV~\cite{PhysRevB.54.14153} for intrarow H diffusion on the (001)
surface and 1.17 eV (intra row) and 1.49 eV (inter row) on the
Si(110) surface\cite{C1CP20108E}. The diffusion from the top of the
pillar (the (001) surface) to the side of the pillar (the (110)
surface) has a comparable barrier to diffusion on the Si(001) surface,
while diffusion in the opposite direction has a significantly larger
barrier. 
\section{Discussion and Conclusions}

We have studied the stability of hydrogen adatoms at positions
near the intersection of the Si(001) and (110) surfaces, as might be found
on a nanopillar grown on Si(001) by patterned ALE.  We have identified a diffusion pathway around the corner and calculated the barrier energies on it.  We find that that the hydrogen is more stable on the (110) face than the (001) face.

In the PALE context we are concerned with whether hydrogen will leave the top
of the nanopillar for the sidewall or vice versa. The process involves a disilane gas source and leaves hydrogen on the  sidewalls. Hydrogen migration onto the growth surface would interfere with subsequent STM lithography and possibly compromise the entire process.  From the barriers we can see that, at least for this configuration, diffusion off the top of a nanopillar will only be activated once diffusion on the substrate is activated and diffusion back onto the top will occur only at higher temperatures.  The difference in barrier energy ($\approx$ 0.3 eV) is significant in DFT terms and equivalent to a temperature increase of $\approx$ 100 K.

The actual diffusion rate $\nu$ ($s^{-1}$) can be estimated from the energy barrier by the Arrhenius equation 

\begin{equation}
\nu =\nu _{hop}\times \exp \left[ -\left( \frac{\Delta \mathrm{E}}{k_{B}T}%
\right) \right], 
\end{equation}

where $\nu _{hop}$ is the
attempt frequency,  $\Delta \mathrm{E}$ the energy barrier, $k_{B}$ the Boltzmann constant and $T$ the ambient temperture.  $\nu _{hop}$\ is generally found to lie in the
range $10^{12}-10^{13}$ $s^{-1}$ and since the rate expression is
dominated the negative exponential energy term we can take $\nu
_{hop}=10^{13}$ $s^{-1}$ to get an upper bound on the rate. If the PALE process
temperature of 550 K is assumed and adsorption sites are assumed to be
occupied with a probability of one, then we could estimate a rate of
$1.7\times 10^{-3}$ $s^{-1}$ or approximately one diffusion event
every 10  minutes \emph{off} the nanopillar.  The reverse process would be
three hundred times less frequent at 550K. 

We can extrapolate from this single event to an entire nanopillar, under the conditions used in PALE.  A pillar with a side of 5 nm would have 100 edge sites available, generating a diffusion event \emph{off} the pillar every 5  or 6
seconds, assuming that there was a empty site to reach.  The reverse process produces a diffusion event every
half hour.  These results indicate that there
would be net hydrogen migration off the pillar, which is a desirable
outcome in PALE terms.  There is little reason to be
concerned about higher temperatures, as the process temperature of 550K is
chosen to avoid desorption of the hydrogen resist (it begins to show mobility at temperatures exceeding 600K).

Although sidewall hydrogen diffusion may not be a critical issue in
a PALE manufacturing process other obstacles remain.  Chief amongst these is the the formation
of anti-phase boundaries (APBs) in the growth surface due to the collision of islands with different registry. These APBs can trap hydrogen beneath the surface, leading to a reduction in the depassivation yield from subsequent STM lithography.  The reduction in yield can be mitigated by adjusting the STM parameters prior to each depassivation
step, but the remanent hydrogen causes cumulative surface damage which halts
epitaxial growth after 2 or 3 monolayers.  The remediation of
APBs and improvement of the quality of the silicon is the subject of ongoing research.
   
\section{ACKNOWLEDGEMENTS}
We acknowledge useful discussions with James Owen and John Randall of
ZyvexLabs.
\bigskip

\bibliographystyle{jpcm}
\bibliography{HDiffCorner}
\pagebreak

\end{document}